\documentstyle[twocolumn,aps,prl,epsf,floats,epsfig]{revtex}
\newcommand{\beq}{\begin{equation}}
\newcommand{\eeq}{\end{equation}}
\newcommand{\bqa}{\begin{eqnarray}}
\newcommand{\eqa}{\end{eqnarray}}

\def\square{\vcenter{\vbox{\hrule height.4pt
          \hbox{\vrule width.4pt height8pt
          \kern8pt\vrule width.4pt}\hrule height.4pt}}}

\def\sumint{\hbox{$\sum$}\!\!\!\!\!\!\int}

\begin{document}
\preprint{
\vbox{\halign{&##\hfil\cr
        & hep-ph/0106045 \cr
&\today\cr }}}

\title{Solution to the 3-loop $\Phi$-derivable Approximation\\
        for Scalar Thermodynamics}

\author{Eric Braaten and Emmanuel Petitgirard}
\address{Physics Department, Ohio State University, Columbus OH 43210,
USA}

\maketitle

\begin{abstract}
{\footnotesize
We solve the 3-loop $\Phi$-derivable approximation
to the thermodynamics of the massless $\phi^4$ field theory
by reducing it to a 1-parameter variational problem. 
The thermodynamic potential is expanded 
in powers of $g^2$ and $m/T$, where $g$ is the coupling constant,
$m$ is a variational mass parameter, and $T$ is the temperature.  
There are ultraviolet divergences beginning at 6th order in $g$
that cannot be removed by renormalization.
However the finite thermodynamic potential obtained
by truncating after terms of 5th order in $g$ and $m/T$
defines a stable approximation 
to the thermodynamic functions. }
\end{abstract}

\setcounter{page}{1}
\vskip 0.5cm

%\begin{center}
%%%%%%%%%%%%%%%%%%%%%%%%%%%%%%%%%%%%%%%%%%%%%%%%%%%%%%%%%%%%%%%%%%
%{\bf Introduction}
%%%%%%%%%%%%%%%%%%%%%%%%%%%%%%%%%%%%%%%%%%%%%%%%%%%%%%%%%%%%%%%%%%
%\end{center}

The thermodynamic functions for massless relativistic field theories 
at high temperature $T$ can be calculated as weak-coupling expansions
in the coupling constant $g$.
They have been calculated explicitly through order $g^5$ 
for the massless $\phi^4$ 
field theory \cite{Parwani-Singh,Braaten-Nieto:scalar}, for QED 
\cite{Parwani,Andersen}, and for QCD \cite{Zhai,Braaten-Nieto:QCD}.
Unless the coupling constant is tiny, 
the weak-coupling expansions are poorly convergent and
sensitive to the renormalization scale.
This makes the weak-coupling expansion
essentially useless as a quantitative tool:  
it seems to be reliable only when the coupling constant
is so small that the corrections to ideal gas behavior are 
negligibly small.
The physical origin of the instability seems to be
effects associated with screening and quasiparticles. 

A possible solution to this instability problem is to reorganize 
the weak-coupling expansion within a variational framework.  
A variational approximation can be defined by  
a {\it thermodynamic potential} $\Omega$ 
that depends on a set of variational parameters $m_i$.
The free energy and other thermodynamic functions 
are given by the values of $\Omega$ and its derivatives 
at the variational point where $\partial \Omega/\partial m_i = 0$.
A variational approximation is systematically improvable 
if there is a sequence of successive approximations to $\Omega$ 
that reproduce the weak-coupling expansions of the thermodynamic 
functions to successively higher orders in $g$.
An example of a systematically improvable variational approximation
is {\it screened perturbation theory}, which involves a single 
variational mass parameter \cite{K-P-P}.

A variational approach can be useful only if the correct physics 
can be captured by appropriate choices of the variational parameters.
Information about screening and quasiparticle effects is contained
within the exact propagator of the field theory.
The possibility that these effects are responsible for the instability 
of the weak-coupling expansion suggests the use of the 
propagator as a variational function. 
Such a variational formulation was constructed 
for nonrelativistic fermions by Luttinger and 
Ward\cite{Luttinger-Ward} and by Baym \cite{Baym} long ago.
In the case of a relativistic scalar field theory,
the propagator has the form $[P^2 + \Pi(P)]^{-1}$,
where $\Pi(P)$ is the self-energy which depends on the momentum $P$.
The thermodynamic potential has the form
\begin{equation}
\Omega_0[\Pi] =
{1 \over 2} 
\sumint_P \left[ \log \left( P^2 + \Pi \right)
	- {\Pi \over P^2 + \Pi} \right]
+ \Phi [\Pi] \,,
\label{thpot}
\end{equation}
where the interaction functional $\Phi[\Pi]$ can be expressed as a sum of 
2-particle-irreducible diagrams.
It is constructed so that the solution to the variational 
equation $\delta \Omega_0/\delta \Pi = 0$ is the exact self-energy,
and the value of $\Omega_0$ at the variational point 
is the exact free energy.
We can obtain a systematically improvable variational approximation
by truncating $\Phi$ at $n$'th order in the loop expansion, 
where $n = 2,3,...$.
We refer to such an approximation 
as the $n$-loop {\it $\Phi$-derivable approximation}.
The 2-loop $\Phi$-derivable approximation for QCD has recently been used 
as the basis for quasiparticle models of the thermodynamics 
of the quark-gluon plasma \cite{B-I-R,Peshier}.
Since $\Phi$-derivable approximations guarantee consistency with 
conservation laws, they may be particularly useful for nonequilibrium
problems \cite{Knoll}.

While the $\Phi$-derivable approximation is easily formulated,
it is not so easy to solve.  The variational equation is a 
nontrivial integral equation that, to the best of our knowlege,
has never been solved for a relativistic field theory except
in trivial cases where the self-energy is independent of the momentum.
The main problem is that the thermodynamic potential has severe 
ultraviolet divergences that vanish at the variational point 
only if $\Phi$ is calculated to all orders.
They do not vanish away from the variational point, 
and they do not vanish at the variational point 
if the loop expansion for $\Phi$ is truncated.

In this Letter, we solve the 3-loop $\Phi$-derivable
approximation for a massless scalar field theory with a $\phi^4$
interaction
by reducing it to a 1-parameter variational problem 
involving a mass parameter $m$.
The resulting thermodynamic functions have ultraviolet divergences 
beginning at 6th order in the coupling constant $g$.
However the finite thermodynamic potential obtained
by truncating after terms of 5th order in $g$ and $m/T$ 
defines a stable approximation to the thermodynamic functions.

%\begin{center}
%%%%%%%%%%%%%%%%%%%%%%%%%%%%%%%%%%%%%%%%%%%%%%%%%%%%%%%%%%%%%%%%%%
%{\bf Definition of the theory}
%%%%%%%%%%%%%%%%%%%%%%%%%%%%%%%%%%%%%%%%%%%%%%%%%%%%%%%%%%%%%%%%%%
%\end{center}

The lagrangian for the massless scalar field theory 
with a $\phi^4$ interaction with bare coupling constant
$\alpha_0 = g_0^2/(4 \pi)^2$ is 
\bqa
{\cal L} \;=\; \mbox{$1\over2$} \partial_{\mu}\phi\partial^{\mu}\phi
- \mbox{$1\over 24$} g_0^2 \phi^4 \;.
\label{barel}
\eqa
We define the renormalized coupling constant 
$\alpha = g^2/(4 \pi)^2$ using dimensional regularization 
in $4-2 \epsilon$ dimensions 
and the modified minimal subtraction ($\overline{\rm MS}$) 
prescription with renormalization scale $\mu$: 
\bqa
\alpha_0 \mu^{-2 \epsilon} &=&
\alpha  + {3 \over 2 \epsilon} \alpha^2
+ \left( {9 \over 4 \epsilon^2} -  {17 \over 12 \epsilon} \right) 
\alpha^3 + ...  \;.
\label{bare-ren}
\eqa
The two-loop beta function for the running coupling constant 
$\alpha(\mu)$ is $\beta(\alpha) = 3 \alpha^2 - {17\over3} \alpha^3$.
The thermodynamic functions for this theory are known to order $g^5$
\cite{Parwani-Singh,Braaten-Nieto:scalar}.
The weak-coupling expansion for the free energy density is
\bqa
{{\cal F} \over {\cal F}_{\rm ideal}} &=& 
1 + 15 \left[ - {\alpha \over 12} 
+ 4 \left( {\alpha \over 6} \right)^{3/2} 
+ 9 (L + 1.097) \left( {\alpha \over 6} \right)^2 \right.
\nonumber
\\ 
&& \left. 
+ 36 \left( 2 \log(\alpha/6) - 3 L + 3.665 \right) 
	\left( {\alpha \over 6} \right)^{5/2} \right] \;,
\label{F-weak}
\eqa
where $\alpha=\alpha(\mu)$, $L = \log(\mu/4\pi T)$, 
and ${\cal F}_{\rm ideal}$ is the pressure of the 
ideal gas of a free massless boson:
${\cal F}_{\rm ideal} = - (\pi^2/90) T^4$.

The instability of the weak-coupling expansion is 
illustrated in Fig.~1, which shows the free energy 
divided by that of the ideal gas as a function of $g(2\pi T)$.
The dashed lines are the predictions of the weak coupling 
expansion (\ref{F-weak}) with $\mu = 2 \pi T$
truncated after orders $g^n$ for $ n = 2,3,4,5$.
The dashed line for $n=2$ is hidden under the solid line labelled $g^2$.
The successive approximations show no sign of converging.

%%%%%%%%%%%%%%%%%%%%%%%%%%%%%%%%%%%%%%%%%%%%%%%%%%%%%%%%%%%%%%%%%%%
% Figure 1
%%%%%%%%%%%%%%%%%%%%%%%%%%%%%%%%%%%%%%%%%%%%%%%%%%%%%%%%%%%%%%%%%%%
\begin{figure}[t]
\vspace*{-1cm}
\epsfig{file=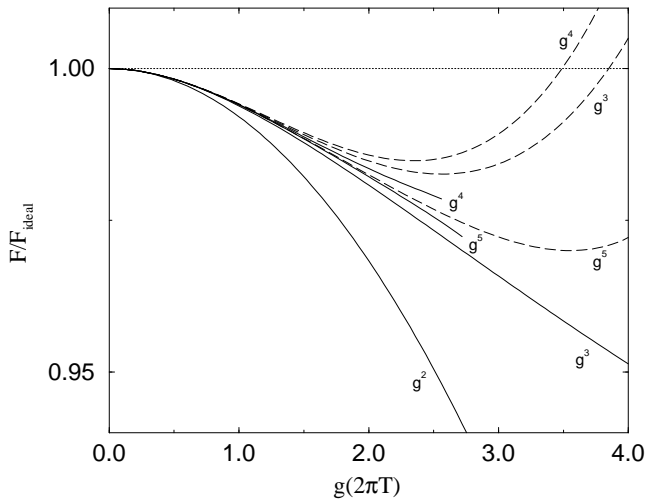,width=7.5cm,angle=-90}
\vspace*{0.5cm}
\caption{Free energy divided by that of the ideal gas
for the weak-coupling expansion (dashed lines) and 
the truncated $\Phi$-derivable approximation (solid lines).}
\end{figure}
%%%%%%%%%%%%%%%%%%%%%%%%%%%%%%%%%%%%%%%%%%%%%%%%%%%%%%%%%%%%%%%%%
%\begin{center}
%%%%%%%%%%%%%%%%%%%%%%%%%%%%%%%%%%%%%%%%%%%%%%%%%%%%%%%%%%%%%%%%%%
%{\bf 2-loop $\Phi$-derivable approximation}
%%%%%%%%%%%%%%%%%%%%%%%%%%%%%%%%%%%%%%%%%%%%%%%%%%%%%%%%%%%%%%%%%%
%\end{center}

To illustrate our method in the simplest possible context,
we first consider the 2-loop $\Phi$-derivable approximation.
The self-energy is independent of $P$, so we denote it by $\Pi = m^2$.
The interaction functional in (\ref{thpot}) is
\begin{eqnarray}
&& \Phi(m) = 
\mbox{$1 \over 8$} g_0^2 \mu^{-4\epsilon} {\cal I}_{\rm tad}^2 ,
\label{phi0}
\\
&& {\cal I}_{\rm tad} =\sumint_Q {1 \over Q^2 + m^2}  . 
\nonumber
\end{eqnarray}
The measure  for the dimensionally regularized 
sum-integrals in (\ref{thpot}) and (\ref{phi0})
includes a factor of $(e^\gamma \mu^2/4 \pi)^{\epsilon}$, 
where $\gamma$ is Euler's constant.
We insert a factor of $\mu^{-2 \epsilon}$
into the sum-integral in (\ref{thpot}), 
so $\Omega_0$ is independent of the
renormalization scale.
The variational equation $\partial \Omega_0/\partial m = 0$
reduces to the simple gap equation
\begin{equation}
m^2 = \mbox{$1 \over 2$} g_0^2 \mu^{-2\epsilon} {\cal I}_{\rm tad} .
\label{gap2:1}
\end{equation}
The sum-integrals in (\ref{thpot}) and (\ref{phi0})
contain ultraviolet divergences.
Some of the divergences vanish at the variational point.
They can be removed 
by adding a term proportional to the square of the gap equation:
\begin{equation}
\Omega \;=\; \Omega_0
- (m^2 - G)^2/(32 \pi^2 \alpha_0)  \,,
\label{addterm2}
\end{equation}
where $G$ is the expression on the right side of (\ref{gap2:1}).
The additional term gives no contribution to those
thermodynamic functions that are determined by the value of $\Omega$ 
and its first derivatives at the variational point.
The remaining divergences can be absorbed into a renormalized 
coupling constant $\bar \alpha(\mu)$ defined by 
$\alpha_0 \mu^{-2 \epsilon} =
\bar \alpha (1 - \bar \alpha/\epsilon)^{-1}$.
Expanding in powers of $m/T$,
the resulting thermodynamic potential $\Omega$ is
\begin{eqnarray}
\Omega /{\cal F}_{\rm ideal} = 1 
&+& 15 \left[ 3\hat m^4 / \bar \alpha -\hat{m}^2 + 4\hat{m}^3 \right.
\nonumber
\\
&& \left. \hspace{1cm}
	+ 3(L+\gamma) \hat{m}^4 + ... \right],
\label{poth:N6}
\end{eqnarray}
where $\hat m = m/(2 \pi T)$. 
The coupling constant $\bar \alpha$ runs with the beta function
$\beta(\bar \alpha) = \bar \alpha^2$, 
whose first coefficient is too small by a factor of 3 
compared to that of the true coupling constant.
Thus $\bar \alpha$ can be identified with the true coupling constant
$\alpha$ only at a single scale $\mu_0$.
If one expresses the thermodynamic potential in terms of the true 
coupling constant $\alpha$ defined by (\ref{bare-ren}),
the ultraviolet divergences cannot be eliminated.
The closest one can come to defining a finite thermodynamic potential 
is to truncate (\ref{poth:N6}) after the term of order $m^3$:
\begin{equation}
\Omega / {\cal F}_{\rm ideal} = 1 
+ 15 \left[ 3 \hat m^4 / \alpha
	-\hat{m}^2 + 4\hat{m}^3  \right]\,.
\label{poth:lo}
\end{equation}
The gap equation obtained by varying (\ref{poth:lo}) with respect to $m$ 
with $\alpha$ fixed is
\begin{equation}
\hat{m}^2 = \alpha \left(\mbox{$1 \over 6$}  - \hat{m} \right)\,. 
\label{gap:lo}
\end{equation}
Solving this quadratic equation and substituting the solution into 
(\ref{poth:lo}), the free energy is
\begin{eqnarray}
{\cal F} / {\cal F}_{\rm ideal} 
= 1 &-& 5 \alpha / 4
\left[ 1+6 \alpha+6 \alpha^2 \right.
\nonumber
\\
&& \hspace{1cm} \left. -4\left(2+3 \alpha\right)
\sqrt{ \alpha/6+\alpha^2/4 } \right] \,,
\label{free:sqrt} 
\end{eqnarray} 
where $\alpha = \alpha(\mu)$.
When expanded in powers of $g$, this agrees with the 
weak-coupling expansion (\ref{F-weak}) through order $g^3$.
It depends on the renormalization scale $\mu$ 
through the dependence of $\alpha(\mu)$ on $\mu$.
However, it is much less sensitive to variations in $\mu$ 
than the truncated weak-coupling expansion.
The reason is that the truncated 
weak-coupling expansion grows like a power of $\alpha$ in the 
strong-coupling limit, while the expression (\ref{free:sqrt}) 
approaches the limiting value ${31 \over 36}$.

%\begin{center}
%%%%%%%%%%%%%%%%%%%%%%%%%%%%%%%%%%%%%%%%%%%%%%%%%%%%%%%%%%%%%%%%%%
%{\bf 3-loop $\Phi$-derivable approximation}
%%%%%%%%%%%%%%%%%%%%%%%%%%%%%%%%%%%%%%%%%%%%%%%%%%%%%%%%%%%%%%%%%%
%\end{center}

We now proceed to consider the three-loop $\Phi$-derivable approximation.
The interaction term in (\ref{thpot}) is 
\begin{eqnarray}
&&\Phi[\Pi] = 
\mbox{$1 \over 8$} g_0^2 \mu^{-4 \epsilon} \Im _{\rm tad}^2
- \mbox{$1 \over 48$} g_0^4 \mu^{-6\epsilon} \Im _{\rm ball} ,
\label{phi3}
\\
&& \Im_{\rm tad} = \sumint_Q {1 \over Q^2 + \Pi(Q)} ,
\nonumber
\\
&& 
\Im_{\rm ball} =
\sumint_{Q_1 Q_2 Q_3} \prod_{i=1}^4 {1 \over Q_i^2 + \Pi(Q_i)} ,
\nonumber
\end{eqnarray}
where $Q_4= -(Q_1 + Q_2 + Q_3)$ in the definition of the 
basketball sum-integral.  
The variational equation obtained by varying
with respect to $\Pi(P)$ is
\begin{eqnarray}
&&\Pi(P) = 
\mbox{$1 \over 2$} g_0^2 \mu^{-2\epsilon} \Im_{\rm tad} 
- \mbox{$1 \over 6$} g_0^4 \mu^{-4\epsilon} \Im _{\rm sun}(P) \,,
\label{vary3}
\\
&&\Im _{\rm sun}(P) = 
\sumint_{Q_1 Q_2} \prod_{i=1}^3 {1 \over Q_i^2 + \Pi(Q_i)} ,
\nonumber
\end{eqnarray}
where $Q_3= -(P + Q_1 + Q_2)$ in the definition of the sunset
sum-integral.  
The variational equation (\ref{vary3}) is a nontrivial integral equation
for $\Pi(P)$ whose solution is complicated by the presence of severe 
ultraviolet divergences in the sum-integrals. 

Our strategy is to introduce a variational mass parameter $m$ 
that is of order $g T$ in the weak-coupling limit 
and calculate the sum-integrals as double expansions in 
$g$ and $m/T$. 
The variational equation (\ref{vary3}) is then solved for $\Pi(P)$
as a function of $P$ and $m$.
Inserting the solution into (\ref{thpot}) and (\ref{phi3}) and
expanding in powers of $g$ and $m/T$, 
the thermodynamic potential $\Omega$ reduces to 
an algebraic function of $m$.
The final step is to minimize $\Omega$
with respect to the variational parameter $m$.

We define the mass parameter $m$ implicitly by the equation
\begin{eqnarray}
m^2 &=& 
\mbox{$1 \over 2$} g_0^2 \mu^{-2\epsilon} \Im_{\rm tad}
- \mbox{$1 \over 6$} g_0^4 \mu^{-4\epsilon} 
	\Im _{\rm sun}(0,{\bf p})\big|_{p=im}\,.
\label{screen:2}
\end{eqnarray} 
The solution to this gap equation for $m$ 
can be interpreted as the screening mass.  
There are two important momentum scales:  
the {\it hard} scale $2 \pi T$ and the {\it soft} scale $m$. 
The hard region for the momentum $P= (2 \pi nT, {\bf p})$ 
includes $n \neq 0$ for all ${\bf p}$ 
and also $n = 0$ with ${\bf p}$ of order $T$. 
The soft region is $n = 0$ and ${\bf p}$ of order $m$.  
We will solve the variational equation in the two momentum regions
separately.
We expand $\Pi (P)$ in powers of $g_0$ and $m/T$. 
For hard momentum $P$, the expansion has the form
\begin{eqnarray}
\Pi (P) &=& m^2 
\;+\; g_0^4 \mu^{- 4\epsilon} \left[ \Pi_{4,0}(P) +  \Pi_{4,1}(P)
+ \ldots \right]
\nonumber
\\
&& \hspace{0.8cm}
\;+\; g_0^8 \mu^{- 8\epsilon} \left[ \Pi_{8,-2}(P) +  
\ldots \right]
\;+\; \ldots \; \,,
\label{Pi-hard}
\end{eqnarray}
where $\Pi_{n,k}(P)$ is of order $T^2(m/T)^k$ when $P$ is of order $T$.
For soft momentum $P = (0, {\bf p})$, the function $\sigma(p)$
in the self-energy $\Pi (0, {\bf p}) = m^2 + \sigma(p)$ has the form
\begin{eqnarray}
\sigma(p) &=& 
g_0^4 \mu^{- 4\epsilon} \left[ \sigma_{4,-2}(p) + \sigma_{4,0}(p) 
+ \ldots \right]
\nonumber
\\
&& 
\;+\; g_0^8 \mu^{- 8\epsilon} \left[ \sigma_{8,-4}(p) + \ldots \right]
\;+\; \dots \; \,,
\label{Pi-soft}
\end{eqnarray}
where $\sigma_{n,k}(p)$ is of order $m^2(m/T)^k$ when $p$ is of order $m$.
We insert the expansions (\ref{Pi-hard}) and (\ref{Pi-soft}) into the
variational equation (\ref{vary3}) and expand in powers of $g_0^4$ and $m
/T$. 
Matching the coefficients at each order in $g_0^4$ and $m /T$,
we find a recursive structure in which the functions $\Pi_{n,k}(P)$ and 
$\sigma_{n,k}(p)$ are either completely determined or expressed 
in terms of lower order functions.
For example, the solutions to the first few self-energy functions 
at hard momentum $P$ are 
\begin{eqnarray}
\Pi_{4,0} (P)  &=&  
- {1 \over 6} \sumint_{Q_1 Q_2} {1 \over Q_1^2 Q_2^2 Q_3^2} 
	+{1\over 6} T^2 I_{\rm sun} (im)\,,
\label{Pi40}
\\
\Pi_{4,1} (P)  &=& 
- {1 \over 2} T I_1 \sumint_Q \left ( {1 \over Q^2 (P+Q)^2} 
					- {1 \over (Q^2)^2} \right)\,,
\nonumber
\end{eqnarray}
where $Q_3 = -(P+Q_1+Q_2)$.
The solutions to the first few self-energy functions 
at soft momentum $(0,{\bf p})$ are 
\begin{eqnarray}
\sigma_{4,-2} (p)  &=&  
- {1\over 6}T^2 \left[ I_{\rm sun}(p) - I_{\rm sun}(im) \right] \,,
\label{sigma4-2}
\\
\sigma_{4,0} (p)  &=& 
{1 \over 6} (p^2+m^2) 
\sumint_{QR} {Q^2 -(4/d) {\bf q}^2 \over (Q^2)^3 R^2 (Q+R)^2} \,.
\nonumber
\end{eqnarray}
The sunset integral appearing in (\ref{Pi40}) and (\ref{sigma4-2}) is
\begin{eqnarray}
I_{\rm sun}(p) &=& \int_{{\bf q}_1 {\bf q}_2} 
\prod_{i=1}^3 {1\over q_i^2+m^2} \,,
\nonumber
\end{eqnarray}
where ${\bf q}_3 = - ({\bf p} + {\bf q}_1 + {\bf q}_2)$.

Having solved the gap equation, we can reduce the bare
thermodynamic potential $\Omega_0$ to a function of 
the single variational parameter $m$ by
inserting the expansions (\ref{Pi-hard}) and (\ref{Pi-soft})
into (\ref{thpot}) and (\ref{phi3}) and
expanding in powers of $\alpha_0$ and $m/T$.
The function $\Omega_0$ contains ultraviolet 
divergences in the form of poles in $\epsilon$.
Some of them are eliminated when $\Omega_0$ is evaluated at 
the solution of the gap equation. We can cancel them 
through 5th order in $g$ and $m/T$ 
by adding a term proportional to $(m^2 - G)^2$, 
where $G$ is the expression on the right side 
of the gap equation (\ref{screen:2}):
\begin{equation}
\Omega \;=\; \Omega_0 
- \left[ {1 \over \alpha_0}
+ {\Lambda^{-2\epsilon} \over \epsilon}\right]
{(m^2 - G)^2  \over 32 \pi^2}\,.
\label{omega-finite}
\end{equation}
We have introduced an arbitrary momentum scale $\Lambda$
in order that all terms have the correct dimension $4-2 \epsilon$
when $\epsilon \ne 0$.
Other ultraviolet divergences in $\Omega_0$ can be removed 
by using (\ref{bare-ren}) to renormalize the coupling constant. 
There are still other ultraviolet divergences at 6th 
and higher orders in $g$ and $m/T$ that cannot be 
canceled and represent unavoidable ambiguities in the $\Phi$-derivable 
approximation. The most severe divergences 
at 6th order are double poles in
$\epsilon$ proportional to $\alpha m^4$, $\alpha^2 m^2$ and $\alpha^3$.
Expanding (\ref{omega-finite}) in powers of $g$ and $m/T$ 
and then truncating after terms of 5th order, 
we obtain a finite thermodynamic potential:
\begin{eqnarray}
&&\mbox{$1\over 15$} \left( \Omega / {\cal F}_{\rm ideal} - 1 \right) 
\;=\; 
3\hat{m}^4 / \alpha 
\nonumber
\\
&& \hspace{1cm}
+\left[ -\hat{m}^2+4\hat{m}^3+3\left(3L-2\ell+\gamma
\right)\hat{m}^4\right] 
\nonumber
\\
&& \hspace{1cm}
+\alpha\left[ 2\left( \ell-2\log \hat{m} - 2.757642 \right) \hat{m}^2
\right.
\nonumber
\\
&& \hspace{2cm} \left.
-12\left( \ell+\gamma\right)\hat{m}^3\right] 
\nonumber
\\
&& \hspace{1cm}
+\alpha^2\left[ -\mbox{$1\over 6$}\left( \ell-4\log \hat{m}- 9.873296
\right) 
\right. 
\nonumber
\\
&&\hspace{2cm}  \left.
+2\left( \ell-2\log 2+\gamma\right)\hat{m}\right] \,,
\label{omega1}
\end{eqnarray}
where $\alpha = \alpha(\mu)$, $\hat m = m/(2 \pi T)$,
$L=\log(\mu/4 \pi T)$, and $\ell=\log(\Lambda/4 \pi T)$.

The gap equation obtained by varying the finite thermodynamic potential 
(\ref{omega1}) with respect to $m^2$ with $\mu$ and $\Lambda$ fixed is
\begin{eqnarray}
\hat{m}^2 &=& \alpha 
\left[ \mbox{$1\over 6$} - \hat{m} - \left( 3L - 2 \ell + \gamma\right)
\hat{m}^2 \right] \nonumber
\\
&&+\alpha^2
\left[ -\mbox{$1\over 3$}
\left( \ell - 2 \log \hat{m} -3.75764 \right) \right.
\nonumber
\\ && \hspace{1cm} \left.
+3 \left( \ell + \gamma \right) \hat{m} \right]
\nonumber
\\
&&+ \alpha^3
\left[ - \mbox{$1\over 18$} - \mbox{$1\over 6$}
	\left( \ell - 2\log 2 + \gamma \right) \hat m \right]/\hat m^2 \,.
\label{gap:trunc}
\end{eqnarray}
If we solve this gap equation iteratively in powers of $g$  
and insert the solution into (\ref{omega1}),
we recover the weak-coupling expansion (\ref{F-weak}) for the free energy. 
Solving the gap equation numerically, 
we find a solution only below a critical value of $g$
that depends on $\mu$ and $\Lambda$.
For $\mu = \lambda = 2 \pi T$, the critical value is $g = 2.61$.
For larger values of $g$, the thermodynamic potential
(\ref{omega1}) has a run-away minimum at $m = 0$.
A similar behavior was observed in screened perturbation theory 
at 3 loops \cite{A-B-S}.  When the screening mass is used as the 
mass parameter, the solution to the gap equation terminates 
at nearly the same critical value of $g$.

By truncating the thermodynamic potential (\ref{omega1})
and the gap equation (\ref{gap:trunc}) after terms of $n$'th order
in $g$ and $m/T$, where $ n=2,3,4,5$, we obtain a series of 
successive approximations to the free energy.
The stability of these successive approximations is 
exhibited in Fig.~1, which shows the free energy 
divided by that of the ideal gas as a function of $g(2\pi T)$
for $\mu = \Lambda = 2 \pi T$.
For the $g^4$ and $g^5$ truncations,
the solutions to the gap equation terminate at critical values of $g$.
In contrast to the weak-coupling expansion,
the predictions of the truncated $\Phi$-derivable approximation
seem to be converging for values of $g$
below these critical values.
The truncated $\Phi$-derivable approximations
depend on the renormalization scale $\mu$ 
and also for $n=4,5$ on the momentum scale $\Lambda$
introduced in (\ref{omega-finite}).
However the changes in the predictions from varying the 
arbitrary scales is significantly smaller than in the 
weak-coupling expansion.

%\begin{center}
%%%%%%%%%%%%%%%%%%%%%%%%%%%%%%%%%%%%%%%%%%%%%%%%%%%%%%%%%%%%%%%%%%
%{\bf Conclusions}
%%%%%%%%%%%%%%%%%%%%%%%%%%%%%%%%%%%%%%%%%%%%%%%%%%%%%%%%%%%%%%%%%%
%\end{center}

We have solved the 3-loop $\Phi$-derivable approximation
for the thermodynamics of the $\phi^4$ field theory by reducing it to a 
problem with a single variational parameter $m$.
We constructed a finite thermodynamic potential  
by adding a term proportional to the square 
of the gap equation and truncating after terms of 5th order
in $g$ and $m/T$.  This thermodynamic potential has a minimum 
as a function of $m$ only for $g$ below some critical value.
Below this critical value of $g$,
the successive approximations obtained by truncating the 
thermodynamic potential at orders $n=2,3,4,5$
give predictions for the pressure
that are stable with respect to both the order of truncation
and variations in $\mu$ and $\Lambda$.
The resulting predictions for the pressure are numerically close
to those of screened perturbation theory \cite{A-B-S}.
The advantage of our $\Phi$-derivable approximation is that it
represents an infinite-parameter variational approximation.
Our solution to the $\Phi$-derivable approximation
can in principle be extended to higher orders in the loop expansion.
Thus it provides a systematically improvable approximation
to the thermodynamic functions 
that seems to have very good convergence properties.

This work was supported in part by the U.~S. Department of
Energy Division of High Energy Physics grant DE-FG02-91-ER40690. 
We thank J.O.~Andersen and M.~Strickland for valuable discussions.


\begin{thebibliography}{99}

\bibitem{Parwani-Singh}
R.~Parwani and H.~Singh,
%``The Pressure of hot (g**2 phi**4) theory at order g**5,''
	Phys.\ Rev.\ D {\bf 51}, 4518 (1995).

\bibitem{Braaten-Nieto:scalar}
E.~Braaten and A.~Nieto, Phys. Rev. {\bf D51}, 6990 (1995).

\bibitem{Parwani}
C.~Coriano and R.~R.~Parwani,
%``The Three loop equation of state of QED at high temperature,''
	Phys.\ Rev.\ Lett.\  {\bf 73} 2398 (1994);
R.~R.~Parwani,
%``The Free energy of hot QED at fifth order,''
	Phys.\ Lett.\ B {\bf 334}, 420 (1994)
	[Erratum-ibid.\ B {\bf 342}, 454 (1994)];
R.~R.~Parwani and C.~Coriano,
%``Higher order corrections to the equation of state 
%	of QED at high temperature,''
	Nucl.\ Phys.\ B {\bf 434}, 56 (1995).

\bibitem{Andersen}
J.O. Andersen, Phys. Rev. {\bf D53}, 7286 (1996).

\bibitem{Zhai}
P.~Arnold and C.~Zhai,
%``The Three loop free energy for pure gauge QCD,''
	Phys.\ Rev.\ D {\bf 50}, 7603 (1994);
%``The Three loop free energy for high temperature QED and QCD with
fermions,''
	ibid.\  D {\bf 51}, 1906 (1995);
C.~Zhai and B.~Kastening,
%``The Free energy of hot gauge theories with fermions through g**5,''
	Phys.\ Rev.\ D {\bf 52}, 7232 (1995).

\bibitem{Braaten-Nieto:QCD}
E.~Braaten and A.~Nieto, Phys. Rev. Lett. {\bf 76}, 1417 (1996);
       Phys. Rev. {\bf D53}, 3421 (1996).

\bibitem{K-P-P}
F. Karsch, A. Patk\'os, and P. Petreczky, Phys. Lett. {\bf B401}, 69
(1997).

\bibitem{Luttinger-Ward}
J.M. Luttinger and J.C. Ward, Phys. Rev. D {\bf 118}, 1417 (1960).

\bibitem{Baym}
G. Baym, Phys. Rev. {\bf 127}, 1391 (1962).

\bibitem{B-I-R}
J.~P.~Blaizot, E.~Iancu and A.~Rebhan,
%``The entropy of the QCD plasma,''
Phys.\ Rev.\ Lett.\  {\bf 83}, 2906 (1999);
%``Self-consistent hard-thermal-loop thermodynamics 
%	for the quark-gluon plasma,''
Phys.\ Lett.\ B {\bf 470}, 181 (1999);
%``Approximately self-consistent resummations for the thermodynamics 
%	of  the quark-gluon plasma. I: Entropy and density,''
Phys.\ Rev.\ D {\bf 63}, 065003 (2001).

\bibitem{Peshier}
A.~Peshier,
%``HTL resummation of the thermodynamic potential,''
Phys.\ Rev.\ D {\bf 63}, 105004 (2001).

\bibitem{Knoll}
Y.~B.~Ivanov, J.~Knoll and D.~N.~Voskresensky,
%``Self-consistent approximations to non-equilibrium many-body theory,''
Nucl.\ Phys.\ A {\bf 657}, 413 (1999).

\bibitem{A-B-S}
J.~O.~Andersen, E.~Braaten and M.~Strickland,
%``Screened perturbation theory to three loops,''
Phys.\ Rev.\ D {\bf 63}, 105008 (2001);
J.~O.~Andersen and M.~Strickland,
%``Mass expansions of screened perturbation theory,''
hep-ph/0105214.

\end{thebibliography}
\end{document}